\documentclass[10pt]{iopart}

\usepackage{epsfig}
\usepackage{graphics}
\usepackage{graphicx}
\usepackage{bm}
\usepackage{color}
\usepackage{hyperref}
\usepackage{amssymb}
\usepackage{longtable}
\usepackage{rotating}
\usepackage{color}
\usepackage{epsf}

\def\bea{\begin{eqnarray}}
\def\eea{\end{eqnarray}}
\newcommand{\be}{\begin{equation}}
\newcommand{\ee}{\end{equation}}
\newcommand{\ber}{\begin{eqnarray}}
\newcommand{\eer}{\end{eqnarray}}

\newcommand{\ie}{\emph{i.e.} }
\newcommand{\eg}{\emph{e.g.} }

\def\º{\textrm{\textordmasculine}}

\newcommand{\lambdav}{\vec \lambda}
\newcommand{\lambdavsub}{\vec \lambda}

\def\mHz{\mathrm{mHz}}
\def\nHz{\mathrm{nHz}}
\def\GHz{\mathrm{GHz}}

\def\yr{\mathrm{yr}}
\def\Hz{\mathrm{Hz}}

\def\rad{\mathrm{rad}}

\def\A{\mathcal{A}}
\def\C{\mathcal{C}}
\def\lati{\ell}

\begin{document}

% Title

\title[MCMC searches for Galactic binaries in MLDC-1B data sets]
{Markov chain Monte Carlo searches for Galactic binaries in Mock LISA Data Challenge 1B data sets}

\author{Miquel~Trias$^1$, Alberto~Vecchio$^2$ and John~Veitch$^2$}
\address{$^1$ Departament de F\'{\i}sica, Universitat de les Illes
Balears, Cra. Valldemossa Km. 7.5, E-07122 Palma de Mallorca, Spain}
\address{$^2$ School of Physics and Astronomy, University of Birmingham, Edgbaston, Birmingham B15 2TT, UK}
\eads{\mailto{miquel.trias@uib.es}, \mailto{av@star.sr.bham.ac.uk}, \mailto{jveitch@star.sr.bham.ac.uk}}

%%%%%%%%%%%%%%%%%%%%%%%%%%%%%%%%%%%%%%%%%%%%%%%%%%%%%%%%%%%%%%%
\begin{abstract}
We are developing a Bayesian approach based on Markov chain Monte Carlo techniques
to search for and extract information about white dwarf binary systems with the Laser Interferometer 
Space Antenna (LISA). Here we present results
obtained by applying an initial implementation of this method to some of the data sets released in Round 1B
of the Mock LISA Data Challenges. For Challenges 1B.1.1a and 1b %find the signal, providing
%not only an expected value of each parameter but also $95.5	\%$ confident intervals and posterior PDFs.
the signals were recovered with parameters lying within the $95.5	\%$  posterior probability interval and the correlation
between the true and recovered waveform is in excess of $99 \%$. Results were not submitted for Challenge 1B.1.1c due to some convergence 
problems of the algorithm; despite this, the signal was detected in a search over a $2~\mHz$ band.
%not only an expected value of each parameter but also $95.5	\%$ confident intervals and posterior PDFs.
%Despite we didn't submit results for Challenge 1B.1.1c due to some convergence problems, we also were
%able to find the frequency signal in a wide frequency band ($2~\mHz$) and now we are studying the best way to
%solve that convergence problems.
\end{abstract}
\date{\today}


%%%%%%%%%%%%%%%%%%%%%%%%%%%%%%%%%%%%%%%%%%%%%%%%%%%%%%%%%%%%%%
%\maketitle

\section{Introduction}
\label{sec:intro}

Galactic white dwarf (WD) binary systems are %one of the most promising sources that will be observed by the
guaranteed sources for the 
future Laser Interferometer Space Antenna, LISA \cite{LISA-Pre-Phase}. Despite the simple nature
of the expected gravitational radiation -- a quasi monochromatic signal -- 
%essentially monochromatic signal with possible spin-down or spin-up corrections,
the data analysis task becomes challenging due to the tens of millions of such sources in the Galaxy that radiate in the instrument's observational window, with signals strongly overlapping in time and frequency space at the LISA output 
\cite{Nelemans:2001, Farmer:2003pa, Timpano:2005gm}. As a consequence, LISA is expected to resolve about $10^4$ galactic binaries with the remaining producing a ``confusion noise'', whose exact magnitude will depend on the resolving power of the search methods (as well as the actual astrophysical population in the Galaxy).

Due to the large number of sources to be analysed at the same time and the fact that this number is
unknown, Markov chain Monte Carlo (MCMC) techniques
\cite{Cornish:2005qw, Crowder:2006wh, Crowder:2006eu, Crowder:2007ft, Stroeer:2007tg}
and their extension to Reversible Jump Markov chain
Monte Carlos (RJMCMCs) \cite{Umstatter:2005jd, Stroeer:2006ye, Cornish:2007if} are expected to be one of the most powerful search methods.
There are several instances in which techniques based on Bayesian inference have been successfully implemented in the context of
LISA data analysis, not only in the search for WD binary systems,
but also gravitational radiation from massive-black-hole binary inspirals
\cite{Cornish:2006dt, Cornish:2007jv, Rover:2007iq, Brown:2007smbh, Babak:2008rb, cardiffmbh} and extreme-mass ratio inspirals
\cite{emris_GBPB, emris_cornish}.

In the summer of 2007, the Mock LISA Data Challenge (MLDC) Task Force \cite{Arnaud:2006gm, Arnaud:2006gn}
released a re-issue of the Round 1 challenges, called Challenge 1B~\cite{MLDC1B_report}. %, in
%order to provide new groups the opportunity to develop their codes for LISA data analysis. 
In this paper, we present results obtained by applying an initial implementation of a Bayesian approach
based on a MCMC method to the single galactic binary data sets, Challenge 1B.1.1a-c~\cite{mldc_web}.
Several other groups have tackled this problem using a number of approaches: Cornish \& Crowder \cite{Cornish:2005qw,
Crowder:2006wh, Crowder:2006eu, Crowder:2007ft} developed algorithms based on variations of Monte Carlo Metropolis-Hastings
Samplers and successfully applied them to searches for overlapping sources; MCMC methods for single-source analysis have been
explored by several groups, \eg \cite{Stroeer:2007tg}; Prix \& Whelan \cite{prixwhelan1, prixwhelan2} and Kr\'{o}lak and collaborators
have developed a matched-filtering approach, based on the computation of the $\mathcal{F}$-statistic \cite{JKS}. A summary of the
results obtained in Challenge 1B is provided by the MLDC Task Force \cite{MLDC1B_report}.

All the results that we present here, correspond to the blind challenge
data sets and were obtained before the release of the key files, in December 2007.

%%%%%%%%%%%
%%%%%%%%%%%

\section{Analysis method}

Following a Bayesian approach, we can infer the probability density functions (PDFs) of the vector of the unknown model parameters $\lambdav$, given a data set
$d$ and some prior information $W$, using Bayes' theorem:
\be
p(\lambdav|d,W) = \frac{p(\lambdav|W) ~ p(d|\lambdav,W)}{p(d|W)} \; .
\ee
Here, $p(\lambdav|d,W)$ is the \emph{posterior} probability density function of the parameters given the observed data (\ie it is what we are interested
in), $p(\lambdav|W)$ is the \emph{prior} knowledge we have about the different parameters, $p(d|\lambdav,W)$ is the 
\emph{likelihood function} of the data given the model and finally, $p(d|W)$ is a normalisation factor independent of the unknown model parameters,
and therefore irrelevant to this MCMC analysis.

The gravitational wave (GW) signal emitted by a WD binary system with constant orbital frequency
is characterised by $7$ independent parameters: the frequency of the signal $f$,
its amplitude $\A$, two angles to define the sky location of the source -- here longitude $\phi$ and latitude $\lati$ --
the GW polarisation $\psi$, the inclination angle between the angular momentum of the system and an
unitary vector parallel to the line of sight, $\iota$ and finally, a constant value to fix the initial phase of the signal
$\varphi_0$. In our analysis we also treat the noise level that affects the measurements as unknown; we 
parametrise it with $\sigma^2$ the (constant) variance of the noise contribution in frequency domain in the
small band where the signal lies. As a consequence, the analysis of the data containing a single galactic binary
requires the estimation of an 8-dimensional parameter vector $\lambdav$. 

MCMC methods are well
suited to compute the joint posterior PDF $p(\lambdav|d,W)$, and the {\em marginalised} posterior PDF
for any given parameter (or subset of parameters), 
say $\lambda_1$:
\begin{equation}
p(\lambda_1 | d,W) = 
\int d\lambda_2 \dots \int d\lambda_8\>p(\lambdav | d, W) \,.
\label{e:marg}
\end{equation}
From the PDF above, one can then compute the posterior mean %and the, say, 95\% probability interval defined
as
\begin{equation}
%\fl
\bar{\lambda}_j = \int_{-\infty}^{\infty} 
d \lambda_j\>
\lambda_j p(\lambda_j|d,W) %\,,
%\quad
%\quad 
%0.95 =  \min_{\lambda_{j,{\rm high}} - \lambda_{j,{\rm low}}} \int_{\lambda_{j,{\rm low}}}^{\lambda_{j,{\rm high}}} 
%d \lambda_j\>
%p(\lambda_j|d,W)\,.
%0.95 =  \min_{\delta\lambda_{j}} 
%\int_{\delta\lambda_{j}}
%d \lambda_j\>
%p(\lambda_j|d,W)
\,.
\end{equation}
%
%{\bf the old version is very confusing ( I find) with $y$, $y'$ $i$ $n$ nothing defined); here's what I propose, check if 
%you agree and it's really what's done (the old version is still present commented out}

We have implemented a Metropolis-Hastings MCMC algorithm which has the property
that after an initial ``burn-in'' period (which is discarded in the generation of the PDFs), 
it returns samples
of $\lambdav$ with a probability density equal to the desired posterior $p(\lambdav|d,W)$
\cite{Gamerman_book}.

In an MCMC algorithm, a sequence of points (`a chain') is constructed. The first point is chosen randomly according
to a uniform prior distribution, then subsequent elements of the chain are generated in the following way:
\begin{enumerate}
\item Given the current member of the chain corresponding to the parameter vector $\lambdav$, 
the new member $\lambdav'$ is proposed according to
\be
\lambda_i' = \lambda_i + u~\Delta _i \quad (i=1,\dots, 8)\,,
\ee
where $u$ is a Gaussian random number with zero-mean and unit variance, and $\Delta _i$ sets the
size of the step (see the next Section for more details).

\item The new member is accepted with a probability computed according to the Metropolis-Hastings ratio:
\be
\alpha_{\lambdavsub,\lambdavsub'} = \min\left( 1, \frac{\pi(\lambdav')}{\pi(\lambdav)}\right) \;,
\ee
\end{enumerate}
where $\pi(\lambdav) \propto  p(\lambdav|W) ~ p(d|\lambdav,W)$ is the so-called target distribution. 
%As it is well known, the remarkable property
%of the algorithm is that after an initial ''burn-in'' period (which is discarded in the generation of the PDFs), 
%it returns samples
%of $\lambdav$ with a probability density equal to the desired posterior $p(\lambdav|d,W)$.

%%%%%%%%%%%%%%%%%%%%%%%
%We have implemented an MCMC method based on the Metropolis-Hastings algorithm, which 
%can give marginalised results by simply binning the resulting samples into the desired variables and also
%eliminates the need to explicitly calculate the normalization constant in Bayes' theorem.
%In particular, the algorithm followed to generate the Markov chain is the following:
%
%\begin{enumerate}
%\item Propose the new parameters of the chain following a Gaussian probability distribution
%centred in the old values:
%
%\be
%y_i^{(n+1)} = y_i^{(n)} + u~\Delta _i^{(n)} \; ,
%\ee
%
%where $u$ is a Gaussian random number with zero mean and unitary variance.
%
%\item Accept that proposition with a probability given by the so called Metropolis-Hastings ratio:
%
%\be \hspace{-1cm}
%\alpha_{y,y'} = \min\left( 1, \frac{\pi(y')}{\pi(y)}\right) \; , ~ \mathrm{where} ~~ \pi(y) =  p(m|W) ~ p(d|m,W) \; .
%\ee
%
%\end{enumerate}
%
%The remarkable property of this algorithm is that after an initial burn-in period (which is discarded), it generates samples
%of $y$ with a probability density equal to the desired posterior $p(m|d,W)$.

%%%

Since the signal from a galactic binary is nearly monochromatic at the LISA output, it is advantageous to work in the Fourier domain,
and to consider only the small frequency band where the signal power is concentrated. In our implementation we analyse only
the band $B_w = 1.5 \left[ 2 \left(5+4 \pi f_0 \frac{R}{c} \sin\theta \right) f_m + |\dot{f_0}| T_{obs}\right] $ around $f_0$, where $f_m = 1~yr^{-1}$, 
$f_0$ is the signal frequency at a given reference time, $\dot{f_0}$ its time derivative (for the signal model adopted in MLDC-1B $\dot{f_0} = 0$), and 1.5 is a safety factor. We therefore FFT the time series $X$, $Y$ and $Z$ of the unequal arm 
pseudo-Michelson outputs in which the MLDC data are distributed~\cite{mldc_web}, and construct the two noise-orthogonal 
Time Domain Interferometry (TDI) outputs
\cite{Prince:2002hp}:
\be
A = \frac{2X-Y-Z}{3}
\quad ,
\quad \quad
E = \frac{Z-Y}{\sqrt{3}} \; . \nonumber
\ee
$A$ and $E$ for a given choice of the source parameters are then computed directly in the Fourier
domain using an approximation and software implementation provided by Cornish and Littenberg \cite{Cornish:2007if} 
in which the LISA response function for nearly monochromatic signals is separated
in a \emph{fast} and a \emph{slow} part. The \emph{fast} part is computed analytically in the frequency domain, while
the \emph{slow} one can be evaluated efficiently by sampling it at a much slower rate in the time domain and then doing an FFT.
It takes approximately $10^{-2}$ s, depending on the frequency bandwidth, to generate a single signal in the relevant range, and in
order to avoid spurious oscillations at the edges of the frequency band due to windowing, we generate the signal in a 
band $3/2$ times wider than what is needed, $B_w$, and then select only the central part.

The likelihood function that needs to be computed at each step of the MCMC Metropolis-Hastings algorithm is given by
\be
p(d|\lambdav,W) \propto
 \frac{1}{\sigma^N} \exp\left\lbrace -\frac{1}{2 \sigma^2} \sum_{\alpha = 1}^2 \sum_{k=1}^{N} \left| \tilde{d}_{\alpha,k} - \tilde{h}_{\alpha,k}(\lambdav) \right|^2 \right\rbrace \; ,
\ee
where $\alpha$ labels the TDI channel -- $A$ and $E$ --, $k = 1,\dots,N$ the frequency bin, $\tilde d$ the data set and $\tilde h(\lambdav)$ the predicted signal (model) in the Fourier domain.
%for the the  (for a total of $N$$\hat{d}_j$ and $\hat{h}_j$ represent either $A$ or $E$ in the $N$ bins of our frequency band, and their variance
%will provide us an estimation of the noise level, $\sigma_m^2 \propto S_n(f_0)$, assuming that we are working
%in a sufficiently narrow band in which the noise spectral density can be considered as constant.

%%%%%%%%%%%
%%%%%%%%%%%

\section{Results}

\begin{table}
\caption{Information about the Mock LISA Data Challenges 1B.1.1X, corresponding to a single galactic binary signal. Brackets represent the prior range of the two parameters for the different sources.
The data sets are approximately $1~\yr$ long with a cadence of $15$ seconds.}
\begin{center} \small
\begin{tabular}{ccc}
\br
Challenge         &                SNR                  &        frequency (mHz)       \\
\mr
1B.1.1a             &    $\left[ 10~,~20\right] $       &   $\left[ 0.9~,~1.1\right] $   \\
1B.1.1b             &    $\left[ 10~,~20\right] $       &   $\left[ 2.9~,~3.1\right] $   \\
1B.1.1c             &    $\left[ 10~,~20\right] $       &   $\left[ 9~,~11\right] $       \\
\br
\end{tabular}
\end{center}
\label{Tab.MLDC-1B.1.1props}
\end{table}

%
%%%
%

As a test of the initial implementation of the analysis algorithm, we analysed the three data sets of Challenge 1B.1.1,
each of which contained a single galactic binary signal buried in Gaussian and stationary instrument noise. 
In Table~\ref{Tab.MLDC-1B.1.1props} we summarise the main properties of the three
challenge data sets; the main difference amongst them is the actual value and prior range of the signal frequency. In particular, in Challenge 1B.1.1c the frequency prior region was $10$ times wider than for Challenge 1B.1.1a-b.

The algorithm was exactly the same for each analysis and we adopted constant values for the amplitude of the proposals $\Delta_i$ throughout the evolution of the chains. For the angles, the value of $\Delta_i$ was set to one third of the prior range; for the frequency we used $\Delta_f = 3~\yr^{-1} = 95~\nHz$, corresponding to three frequency bins; 
we evolved the noise level in a logarithmic scale with $\Delta_{\log\sigma} = 0.01$, and for the signal's amplitude
we used $\Delta_\A = 3\times 10^{-25}$. These choices were motivated by reasons of efficiency, however for lack of
time no specific tuning of the MCMC code took place. 

Some additional care is required to probe efficiently the prior range of several source parameters. 
The signal model adopted for galactic binaries is characterised by $5$ angular parameters, each of which 
with an intrinsic periodicity: initial phase
$\varphi_0$ and longitude $\phi$ are modulo $2\pi$; latitude $\lati$, polarisation angle $\psi$ and
orientation $\iota$ are modulo $\pi$. We have limited our parameter space in order to work always with angular values
between $0$ and their period. The co-latitude $\theta$, that we use in our search instead of latitude, is taken 
between $0$ and $\pi$, giving a latitude range in the usual interval $\lati \in (-\pi/2 , \pi/2)$. There exists another symmetry concerning the angular parameters, and it is given by the fact that the transformations $\varphi_0 \to \varphi_0 \pm \pi$ or $\psi \to \psi \pm \pi/2$ produce a change of the waveform's global sign. This implies that we can use \eg $\varphi_0 \in (0, 2 \pi)$ as before, but restrict the polarisation angle range to $\psi \in (0, \pi/2)$, 
%[$\psi \in (-\pi/4, \pi/4)$ would also be correct]
since the $(\pi/2, \pi)$ range can be covered by adding or subtracting $\pi$ to the initial phase.

The most crucial parameter for the search for galactic binaries is the frequency, and it is advantageous to analyse 
any given data set in a number of narrow frequency bands. 
For Challenge 1B.1.1a-b, the signals were so strong that their frequency could be approximately determined by simply computing the power spectrum of the data; the MCMC algorithm was then run on a narrow band around that frequency (see Section~\ref{Sec.narrow_fband}). Challenge 1B.1.1c
was more challenging for us because the prior frequency range was 10 times larger and, the signal being at higher frequency,
the Doppler effect spreads the power over many frequency bins without producing clearly identifiable peaks in the spectrum (see
Section~\ref{Sec.wide_fband}). The results of the analyses are discussed in some detail in the next two subsections.

%%%%%%%%%%

%\subsection{Analyzing a narrow frequency band (results submitted to MLDC-1B)}
\subsection{Challenge 1B.1.1a-b}
\label{Sec.narrow_fband}

\begin{figure}
\begin{tabular}{cc}
Challenge 1B.1.1a & Challenge 1B.1.1b  \\
\includegraphics[width=0.48\textwidth]{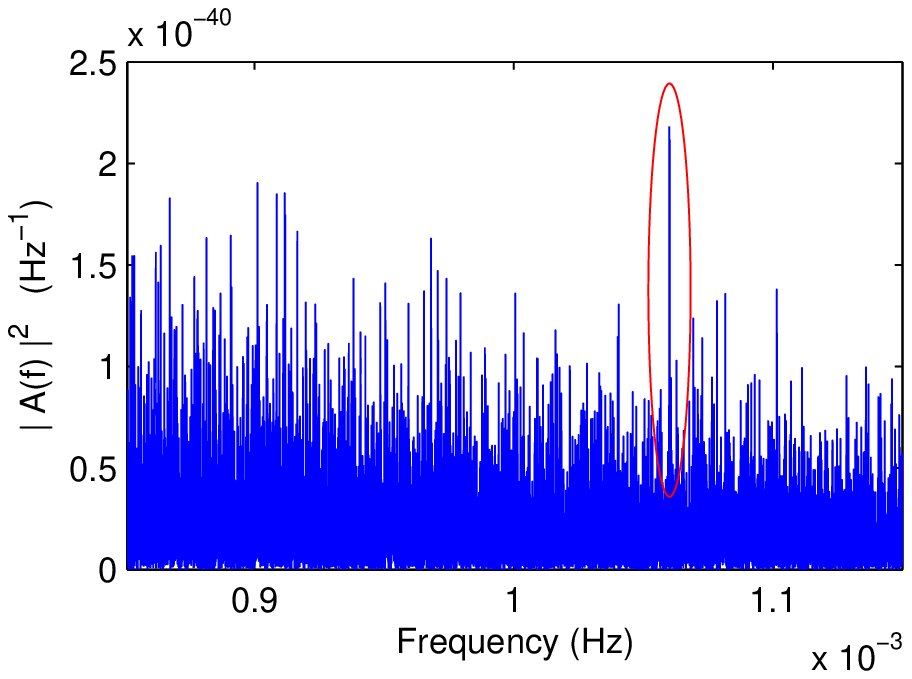} & \includegraphics[width=0.48\textwidth]{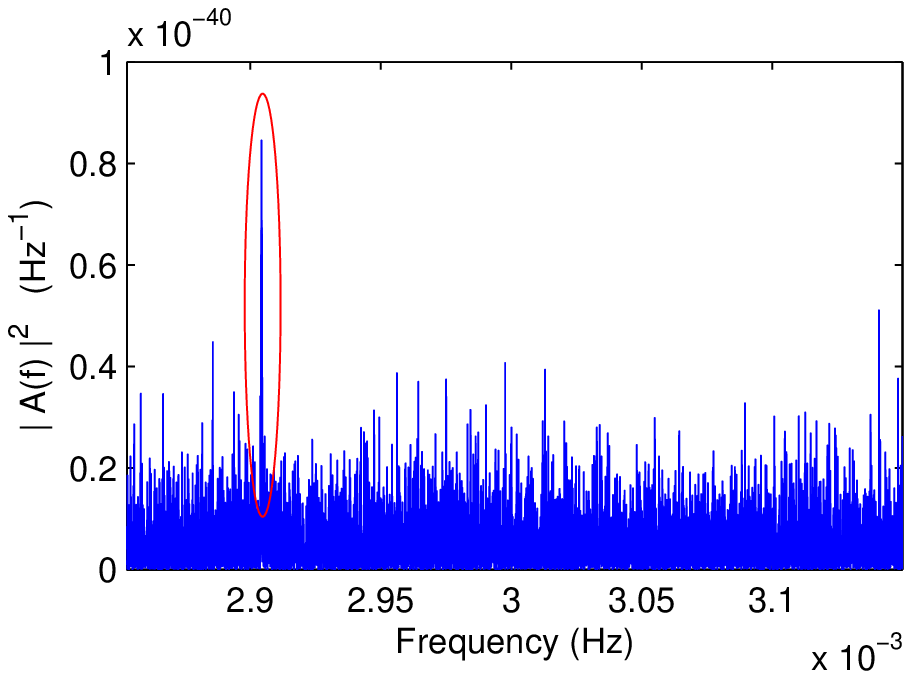}
\end{tabular}
\caption{Power spectral density of LISA's channel $A$ for Challenge 1B.1.1a and 1b. The peak produced by
the actual signal is highlighted by an ellipse.}
\label{Fig.Spectra}
\end{figure}

\begin{figure}
\begin{tabular}{cccc}
\hspace{-0.45cm} \includegraphics[width=0.25\textwidth]{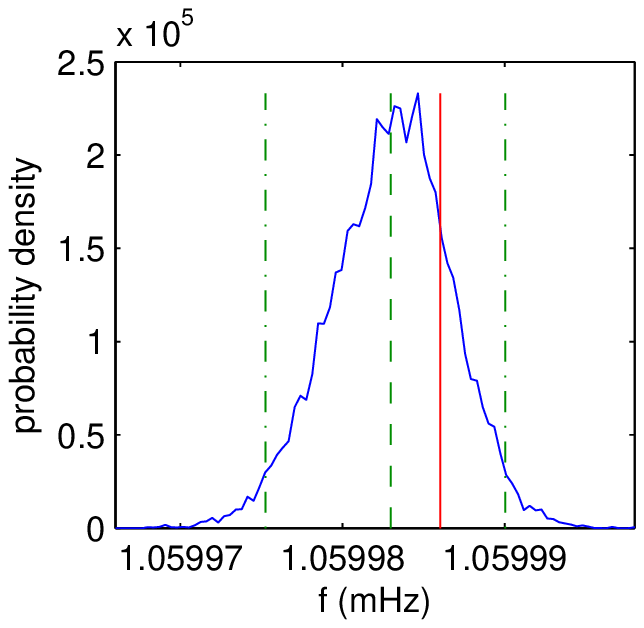} & \hspace{-0.5cm} \includegraphics[width=0.25\textwidth]{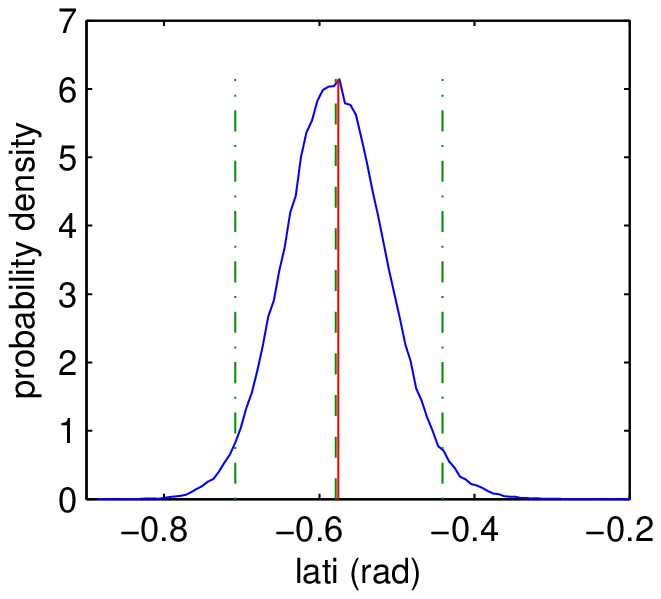} &
\hspace{-0.5cm} \includegraphics[width=0.25\textwidth]{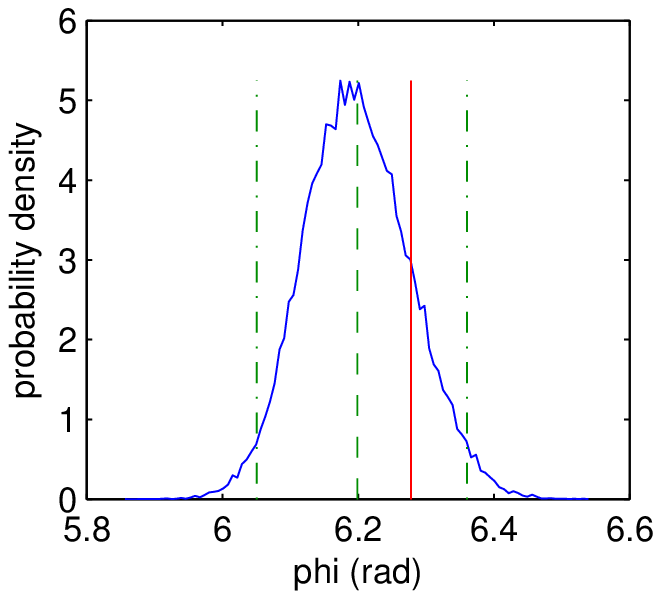} & \hspace{-0.55cm} \includegraphics[width=0.25\textwidth]{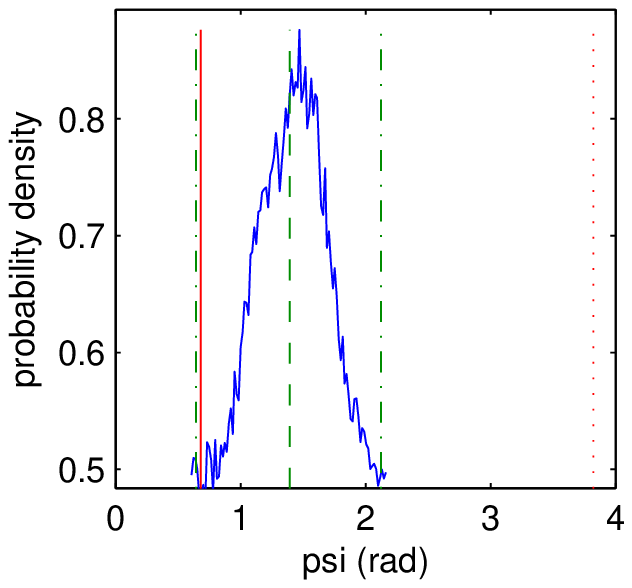} \\
\hspace{-0.45cm} \includegraphics[width=0.25\textwidth]{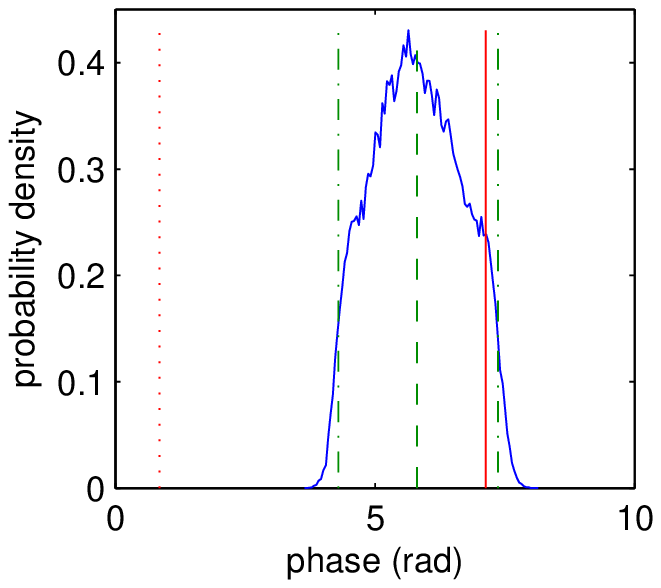} & \hspace{-0.5cm} \includegraphics[width=0.25\textwidth]{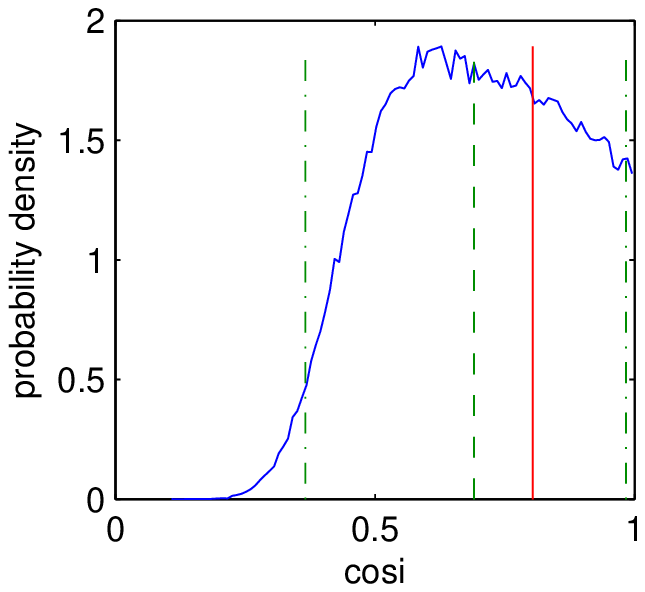} &
\hspace{-0.5cm} \includegraphics[width=0.25\textwidth]{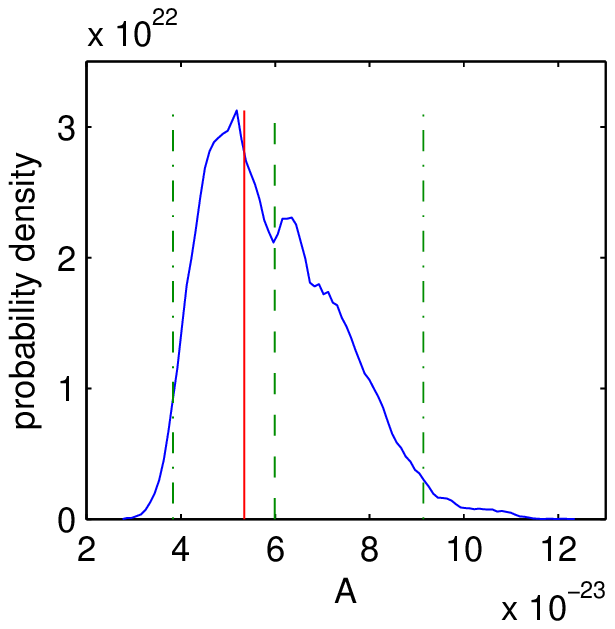} & \hspace{-0.55cm} \includegraphics[width=0.25\textwidth]{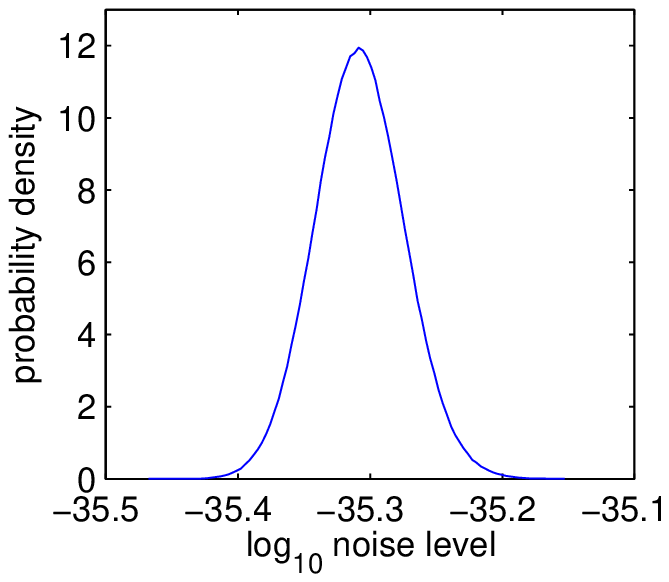}
\end{tabular}
\caption{Marginalised posterior PDFs of each of the seven parameters that characterise the signal,
and noise level for Challenge 1B.1.1a data set. The true values of the parameters are represented by a vertical
solid line, while our submitted values (mean value of the distribution) together with their $95.5\%$ probability interval 
\cite{mldc_howto} are represented by the dashed and dash-dot lines, respectively. Dotted lines indicate positions
in the parameter space equivalent to the true ones.}
\label{Fig.MLDC1B.1.1a}
\end{figure}

\begin{figure}
\begin{tabular}{cccc}
 \hspace{-0.45cm} \includegraphics[width=0.25\textwidth]{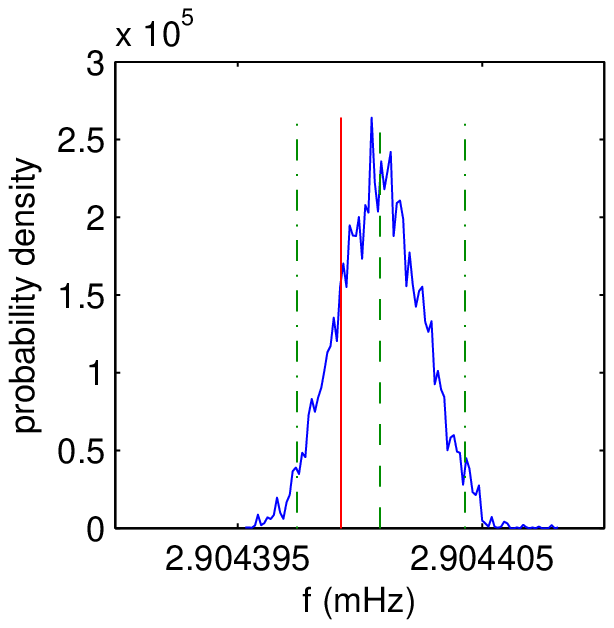} & \hspace{-0.5cm} \includegraphics[width=0.25\textwidth]{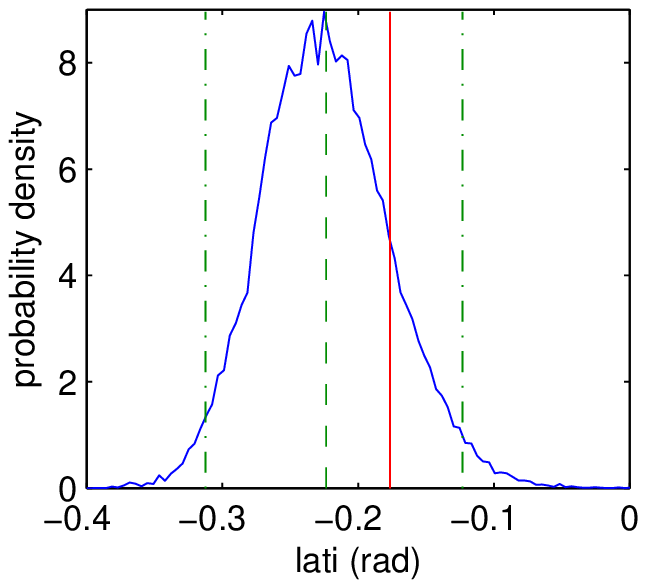} &
 \hspace{-0.5cm} \includegraphics[width=0.25\textwidth]{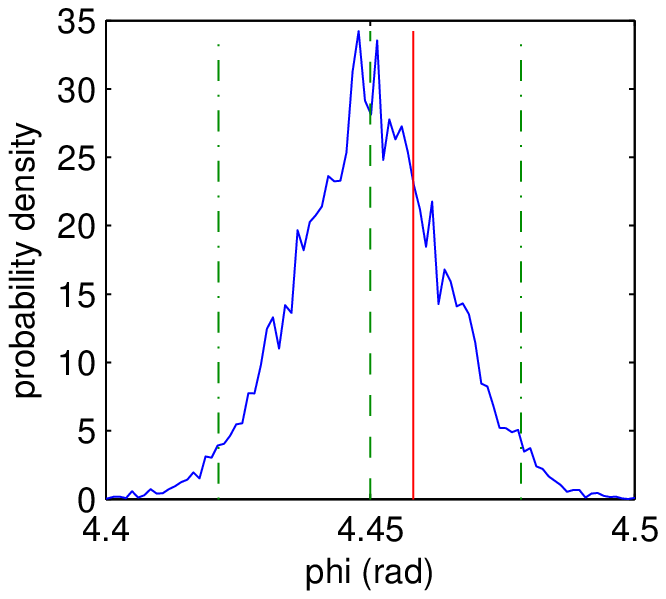} & \hspace{-0.5cm} \includegraphics[width=0.25\textwidth]{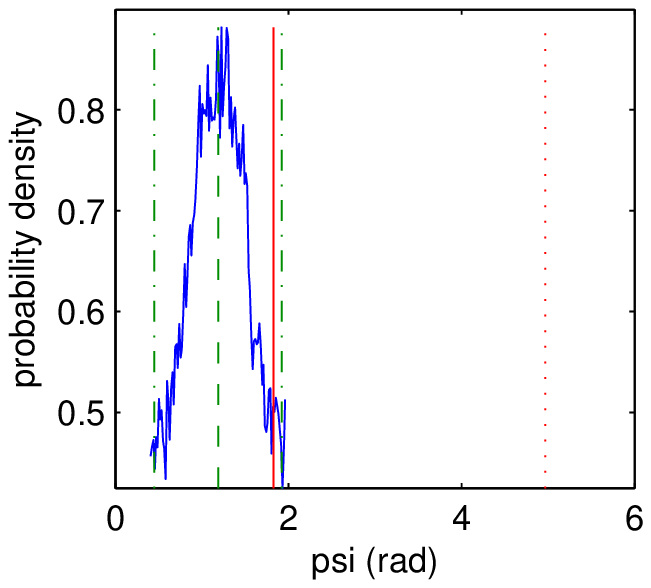} \\
 \hspace{-0.45cm} \includegraphics[width=0.25\textwidth]{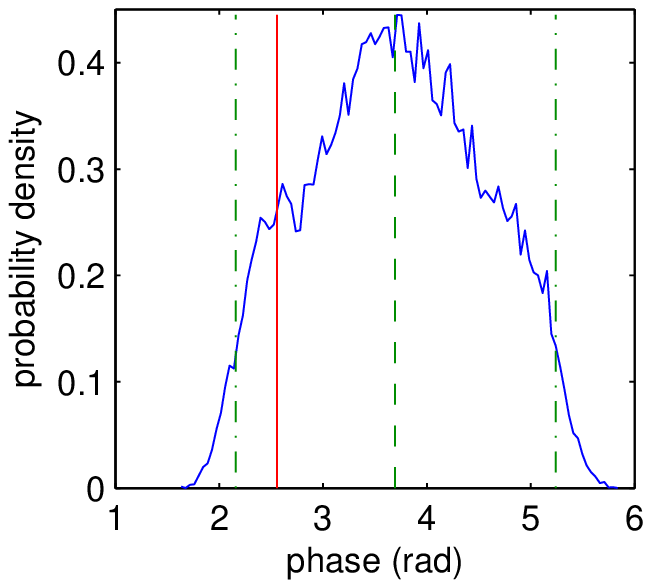} & \hspace{-0.5cm} \includegraphics[width=0.25\textwidth]{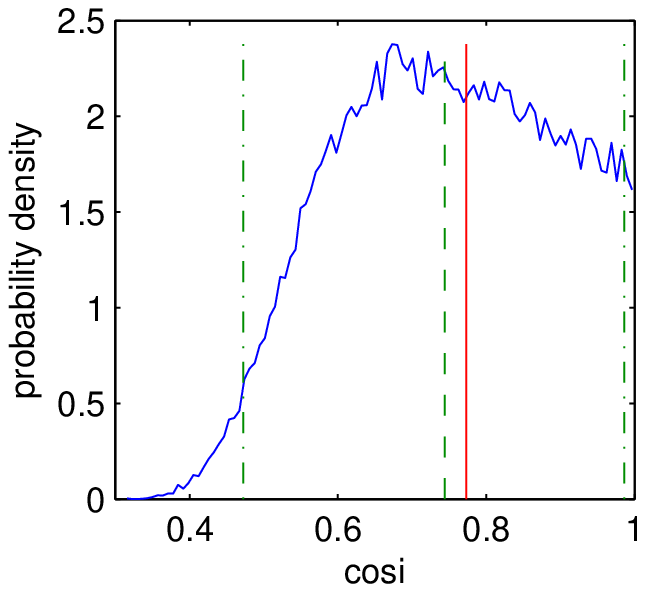} &
 \hspace{-0.5cm} \includegraphics[width=0.25\textwidth]{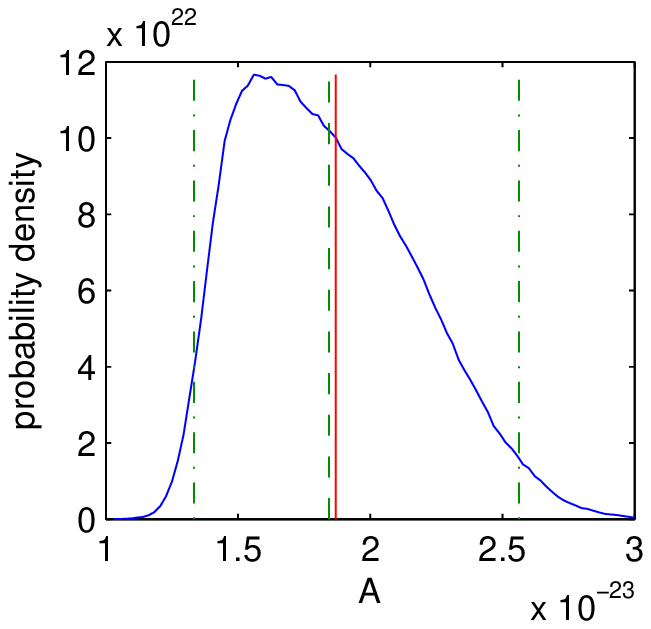} & \hspace{-0.5cm} \includegraphics[width=0.25\textwidth]{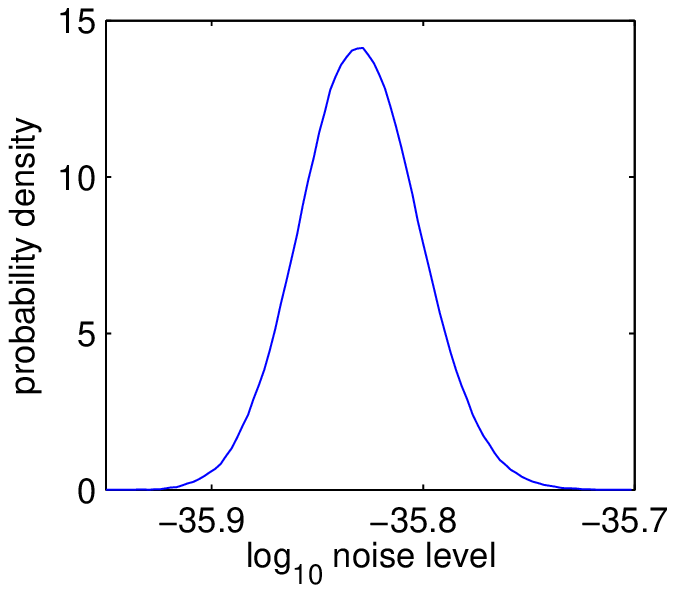}
\end{tabular}
\caption{The same plots as in Figure~\ref{Fig.MLDC1B.1.1a}, but for Challenge 1B.1.1b data set.}
\label{Fig.MLDC1B.1.1b}
\end{figure}

\begin{table}[ht]
\caption{Summary of the results for Challenge 1B.1.1a-b. We show the mean values and the $95.5\%$ probability
interval of the recovered parameters. We also present
the true values from key files, the difference between them and our submitted results [$\Delta \lambda$ (ours)] and
the median of the results submitted by other participating groups \cite{MLDC1B_report} [$\Delta \lambda$ (typ)].
We also show the \emph{optimal} SNR (from the key file), the \emph{recovered} SNR
(values of SNR in $\Delta \lambda$ columns represent
$|\mathrm{SNR}_\mathrm{key} - \mathrm{SNR}_\mathrm{rec}| / \mathrm{SNR}_\mathrm{key}$), as well as
the correlation between the true waveform and the recovered one, $\C$.}
\begin{center} \small
\begin{tabular}{crlccc}
\br
\multicolumn{6}{c}{Challenge 1B.1.1a} \\
Parameter           & \multicolumn{2}{c}{Mean and $95.5\%$ int.}                                                      & True                             & $\Delta \lambda$ (ours)  & $\Delta \lambda$ (typ) \\
\mr
$f$ (mHz)             & $1.0599865$                     & $^{+3.5 \times 10^{-6}}_{-3.9 \times 10^{-6}}$         & $1.0599880$               & $1.51 \times 10^{-6}$ & $1.81 \times 10^{-6}$ \\[3pt]
$\lati$ (rad)               & $-0.57$                             & $^{+0.13}_{-0.14}$                                               & $-0.575$                        & $0.005$                 & $0.017$  \\[3pt]
$\phi$ (rad)          & $6.199$                            & $^{+0.162}_{-0.148}$                                           & $6.278$                         & $0.079$                 & $0.072$  \\[3pt]
$\psi$ (rad)          & $1.39$                               & $^{+0.73}_{-0.74}$                                              & $3.824$                        & $0.708$                  & $0.700$  \\[3pt]
$\varphi_0$ (rad)  & $5.81$                               & $^{+1.56}_{-1.52}$                                              & $0.850$                        & $1.324$                  & $1.320$  \\[3pt]
$\iota$ (rad)          & $0.81$                              & $^{+0.39}_{-0.63}$                                              & $0.637$                        & $0.173$                  & $0.248$  \\[3pt]
$\A \times 10^{23}$ & $5.99$                             & $^{+3.15}_{-2.16}$                                               & $5.34$                         & $0.647$                  & $1.220$  \\[3pt]
SNR                     &  \multicolumn{2}{c}{$13.577$}                                                                          & $13.819$                      &  $0.018$                  & $0.027$  \\[3pt]
$\C$                     &  \multicolumn{3}{c}{ours: $0.992$}                                                                                                           & \multicolumn{2}{c}{typical: $0.988$}  \\
\multicolumn{6}{c}{} \\
\br
\multicolumn{6}{c}{Challenge 1B.1.1b} \\
Parameter           & \multicolumn{2}{c}{Mean and $95.5\%$ int.}                                                      & True                             & $\Delta \lambda$ (ours)  & $\Delta \lambda$ (typ) \\
\mr
$f$ (mHz)             & $2.904401$                      & $^{+3.0 \times 10^{-6}}_{-4.0 \times 10^{-6}}$        & $2.9043992$               & $1.78 \times 10^{-6}$ & $1.92 \times 10^{-6}$  \\[3pt]
$\lati$ (rad)               & $-0.22$                              & $^{+0.10}_{-0.09}$                                               & $-0.176$                           & $0.044$                    & $0.061$  \\[3pt]
$\phi$ (rad)          & $4.45$                               & $^{+0.029}_{-0.029}$                                           & $4.458$                            & $0.008$                    & $0.009$  \\[3pt]
$\psi$ (rad)          & $1.19$                               & $^{+0.73}_{-0.75}$                                              & $4.968$                             & $0.636$                    & $0.736$  \\[3pt]
$\varphi_0$ (rad)  & $3.69$                               & $^{+1.55}_{-1.53}$                                              & $2.556$                             & $1.134$                   & $0.987$   \\[3pt]
$\iota$ (rad)          & $0.73$                              & $^{+0.35}_{-0.56}$                                              & $0.687$                             & $0.043$                    & $0.123$   \\[3pt]
$\A \times 10^{23}$ & $1.84$                             & $^{+0.72}_{-0.51}$                                              & $1.87$                                & $0.029$                   & $0.048$  \\[3pt]
SNR                     &  \multicolumn{2}{c}{$23.479$}                                                                          & $24.629$                            &  $0.047$                  & $0.060$   \\[3pt]
$\C$                     & \multicolumn{3}{c}{ours: $0.996$}                                                                                                                  & \multicolumn{2}{c}{typical: $0.981$}  \\
\br
\end{tabular}
\end{center}
\label{Tab.Submitted}
\end{table}

%
%%%
%

In Figure~\ref{Fig.Spectra} we plot the power spectral density of the $A$ TDI output from Challenge 1B.1.1a-b data sets in the relevant frequency range. It was easy to identify ``by eye'' the approximate frequency of the signal by simply looking at the highest peaks and checking for their presence in the noise orthogonal TDI channel, $E$. Using this procedure, we identified a narrow frequency range ($50$ frequency bins, corresponding to $1.58 \times 10^{-6}~\Hz$) for the analysis with the MCMC code to generate the posterior PDFs of the signal parameters. The typical length of the burn-in stage was chosen to be $2.5 \times 10^4$ steps, and the code run until $10^5$ iterations of the
Markov chain were completed, which in all cases took less than two hours to run on a $2.16~\GHz$ processor. The results submitted for the challenge were obtained by combining the post-burn-in stage of five different runs, each of which used different initial data and seeds for the random number generator.

In Figures~\ref{Fig.MLDC1B.1.1a} and \ref{Fig.MLDC1B.1.1b}, and Table~\ref{Tab.Submitted}, we summarise the results of the analysis. In order to assess the performance of our approach we compare our results with the parameters that describe the actual signal present in the data set -- the ``key'' file -- and also the outcome of the analysis from other participating groups. Our analysis pipeline performed in general in a satisfactory way: the true parameters of the signals are all within the $95.5\%$ posterior probability interval, and the difference between the true and recovered mean value of each parameter is comparable with that obtained by other groups.
Notice that the accuracy of the measurements varies considerably for the seven parameters of the signal: the two sky location angles can be estimated with
an error that is typically of the order of $0.1~\rad$, but the error on the inclination angle is around $0.5~\rad$, and for the polarisation angle and initial phase it increases to $\approx 0.7~\rad$ and $\approx 1.5~\rad$, respectively. The signal frequency can be estimated with very high accuracy, $3.5$ nHz, which represents around one tenth of
a frequency resolution bin of width 32 nHz. Finally, the logarithm of the noise level estimation provided by the MCMC code has a typical error of $0.08$, whereas the
 relative error of the signal's amplitude estimation is typically $35\%$.

Additional quality indicators of the results of a given analysis that are used the context of the MLDCs, are 
the \emph{correlation} $\C$ between the true waveform $h_{\rm key}$ and the recovered one $h_{\rm rec}$, and the 
recovered signal-to-noise ratio $\mathrm{SNR}_\mathrm{rec}$ with respect to the optimal one 
$\mathrm{SNR}_\mathrm{key}$~\cite{MLDC1B_report}. Our results yield $\C$ that differs from 1 by less than $1 \%$, and
$\mathrm{SNR}_\mathrm{rec} = 13.577$ ($\mathrm{SNR}_\mathrm{key} = 13.819$) for Challenge 1B.1.1a and 
$\mathrm{SNR}_\mathrm{rec} = 23.479$ ($\mathrm{SNR}_\mathrm{key} = 24.629$) for Challenge 1B.1.1b. Both
figures of merit are satisfactory and comparable with those obtained by other participating groups, see 
Table~\ref{Tab.Submitted}, and Table 1 of \cite{MLDC1B_report}.

%
%\be
%\C = \frac{\left(h_{\rm key}|h_{\rm rec}\right)}{\sqrt{\left(h_{\rm key}|h_{\rm key}\right)\left(h_{\rm rec}|h_{\rm rec}\right)}} \; .
%\ee
%
%The waveform recovered by our analysis method produces a correlation with that present in the actual data set that differs from 1 by less than 1\%, a high quality figure of merit that is again comparable with that obtained by other participating groups, see Table~\ref{Tab.Submitted}, and Table 1 of \cite{MLDC1B_report}. 

In summary, the outcome of the analysis on the Challenge 1B.1.1a-b data sets provides a successful validation of our analysis pipeline for the simplest possible scenario, and suggests that the inner core of our approach is sound and can be regarded as a solid starting point for more complex searches. However, a realistic analysis of LISA data  needs to deal with tens of thousands of white dwarf binary systems, distributed over a wide ($\sim$ tens of mHz) frequency window and for a range of signal-to-noise ratios. 

Due to the nature of the $A$ parameter, we should expect the distribution of $\log A$ to be approximately symmetrical, whereas we
have sampled and shown $A$ itself, which is asymmetrical. The $\cos (\iota)$ parameter is correlated with $A$, and so its posterior distribution 
is also asymmetrical.

At the time of the deadline for Challenge 1B our analysis approach was still immature and the first natural extension of the code -- \ie the ability to search automatically over a wider frequency band than that required for Challenge 1B.1.1a-b -- was still under development. In fact, results were not submitted for Challenge 1B.1.1c. For completeness, in the next Section we shortly summarise the problems that were encountered and justify the lack of an entry for Challenge 1B.1.1c. This also provides a natural justification for the on-going development work.

%%%%%%%%%%

%\subsection{Searches in a wider frequency band}
\subsection{Challenge 1B.1.1c}
\label{Sec.wide_fband}

\begin{figure}
\begin{tabular}{cc}
(a) training 1B.1.1c & (b) challenge 1B.1.1c \\
\includegraphics[width=0.48\textwidth]{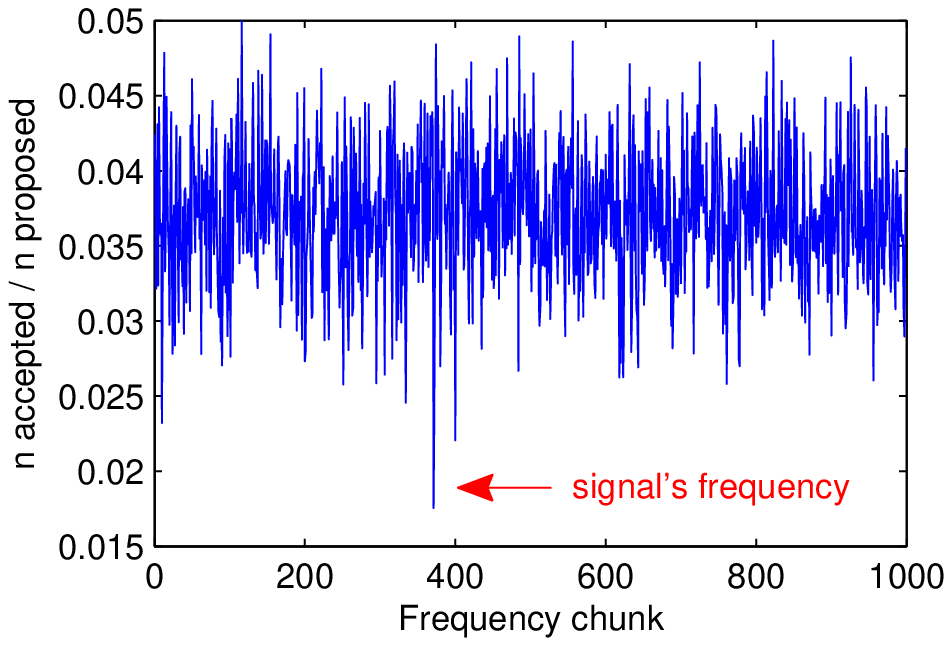} & \includegraphics[width=0.48\textwidth]{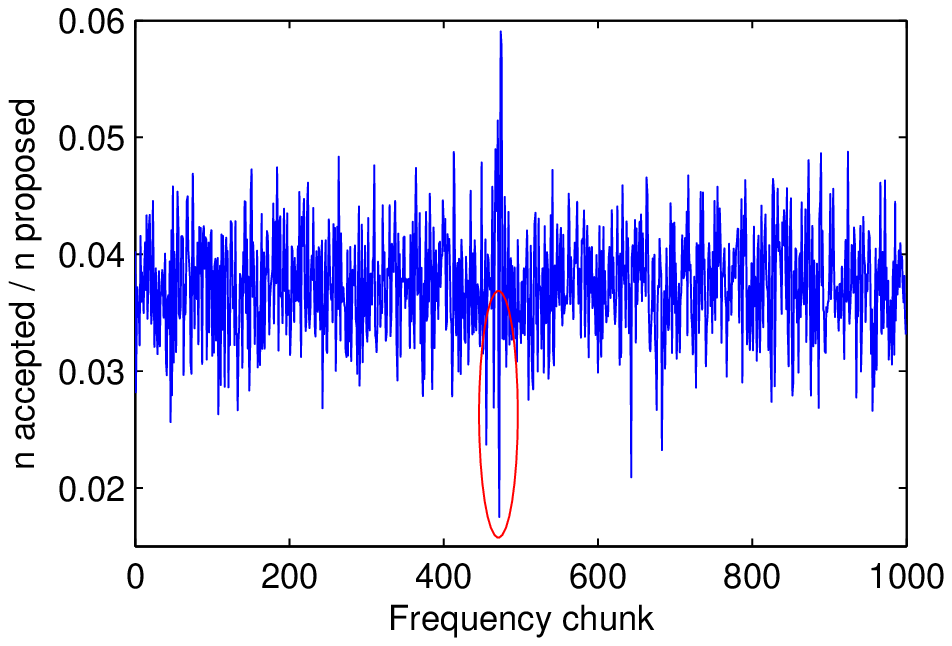}
\end{tabular}
\caption{Looking for the frequency value in Challenge 1B.1.1c, we used the final ratio between the number
of accepted transitions in $f$, over the number of proposed, as a signal presence indicator, expecting to
find a minimum. Here we plot this quantity for the $1000$ different frequency chunks we divided the data into, for
(a) the training data set and (b) the challenge data set. For this latter case, we highlight with an ellipse the absolute minimum of this quantity, expecting to find a signal in that frequency region.}
\label{Fig.MLDC1B.1.1c_effs}
\end{figure}

\begin{figure}
\begin{tabular}{ccc}
(a) & (b) & (c) \\
\includegraphics[width=0.3\textwidth]{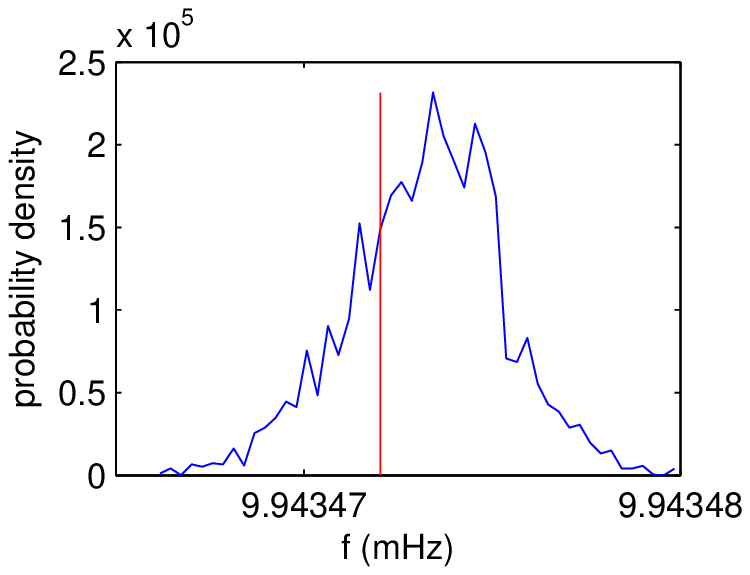} & \includegraphics[width=0.3\textwidth]{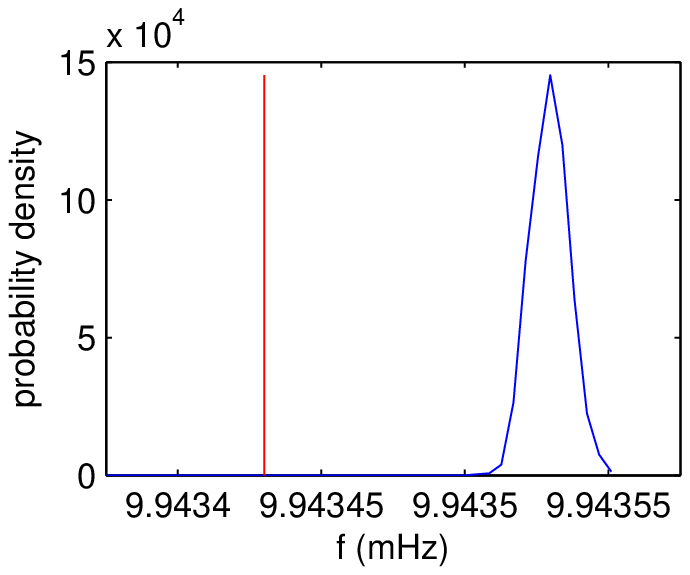}  & \includegraphics[width=0.3\textwidth]{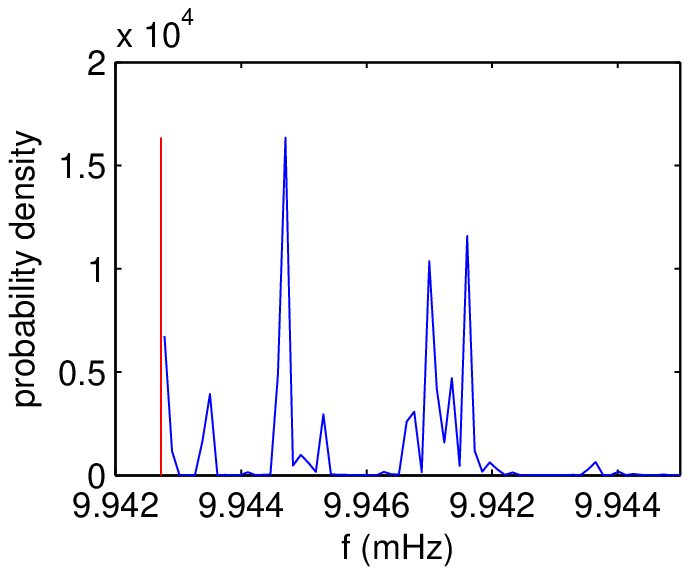} \\
\includegraphics[width=0.3\textwidth]{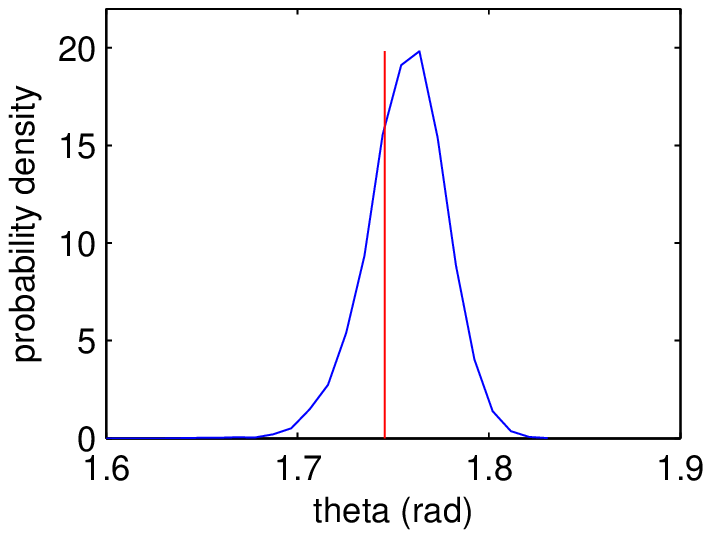} & \includegraphics[width=0.3\textwidth]{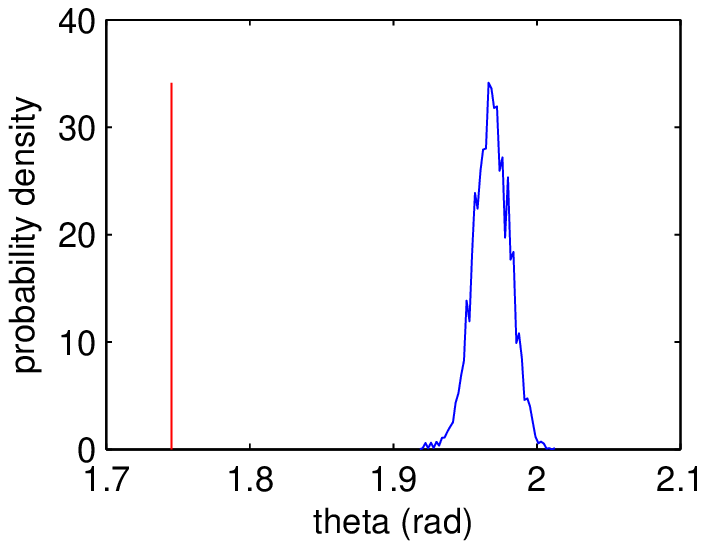}  & \includegraphics[width=0.3\textwidth]{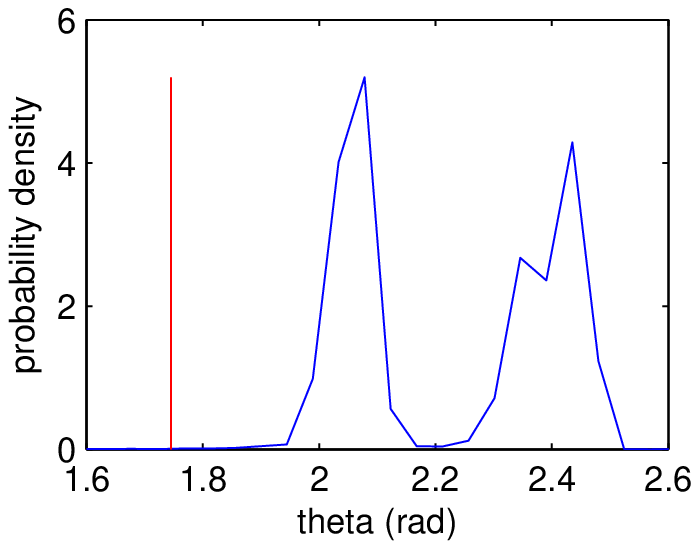} \\
\includegraphics[width=0.3\textwidth]{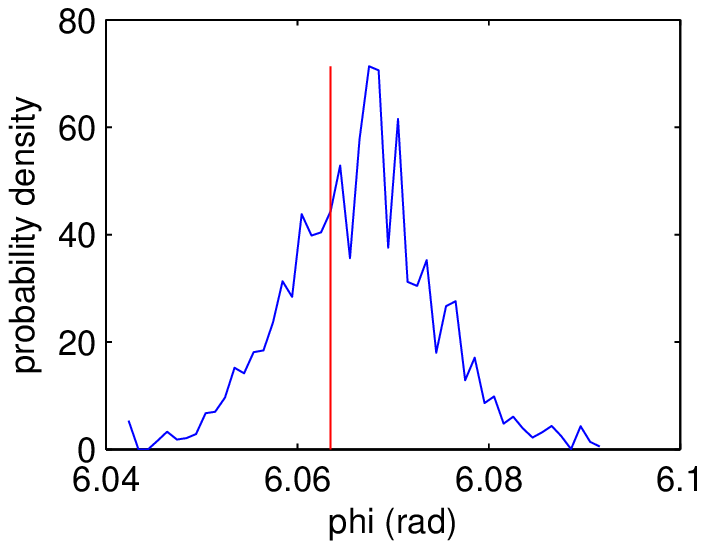} & \includegraphics[width=0.3\textwidth]{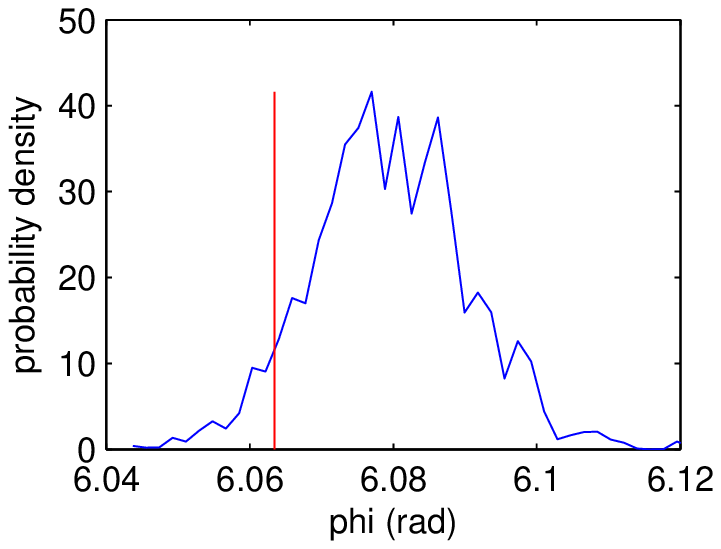}  & \includegraphics[width=0.3\textwidth]{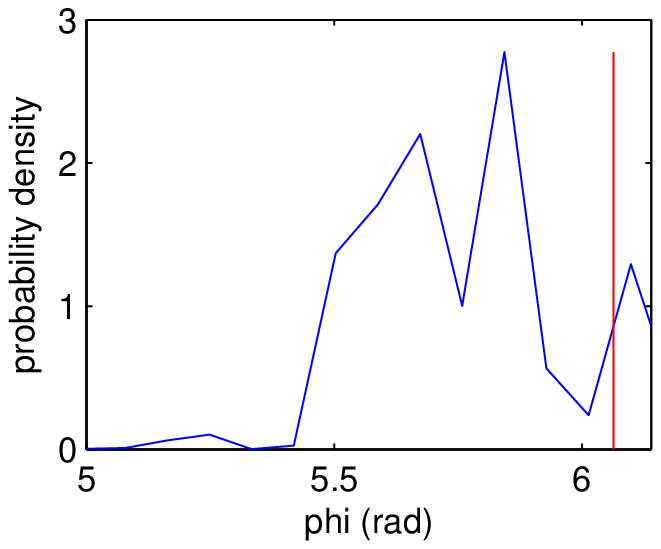}
\end{tabular}
\caption{Marginalised PDFs of frequency and sky location (here, $\theta$ represents the co-latitude) parameters,
after running our MCMC code on the Challenge 1B.1.1c
data set around the frequency value identified from Figure~\ref{Fig.MLDC1B.1.1c_effs}b. Cases (a), (b) and (c) correspond to
independent results obtained just by changing the starting values of the chain and the initialisation of the random number generator.}
\label{Fig.MLDC1B.1.1c}
\end{figure}

%
%%%
%

The efficiency of an MCMC search method for galactic binaries depends crucially on the width of the prior frequency range of the signal. For Challenge 1B.1.1a-b we were able to further restrict the prior range following the \emph{ad hoc} strategy described in the previous Section. This was not possible for Challenge 1B.1.1c, as no clear peak(s) could be identified in the raw power spectrum of the data. This is due to several (simple) reasons: the wider (by a factor 10) prior frequency interval and the higher frequency of the signal, that produces a larger absolute Doppler frequency shift and therefore spreads the signal power over many frequency bins.

In principle one could just let the MCMC code run over a 2 mHz band and eventually the algorithm should converge; however this can take an unreasonably long time and does not provide any real advance in developing an algorithm for more complex and realistic situations. The interim solution that we adopted was to divide the whole frequency band into $1000$ overlapping intervals of 63 frequency bins per interval, and we
run the MCMC code in each of them. As an indicator for the presence of a signal, we used the final ratio between the number of accepted and proposed transitions in $f$, expecting to find a minimum -- the Markov chain has converged and the
transition acceptance probability diminishes -- when a galactic binary signal is indeed
present.

We tested this procedure on the training data set and we found a minimum in the frequency chunk that contained the signal, see Figure~\ref{Fig.MLDC1B.1.1c_effs}a. When running the analysis on the challenge data set, the results contained some additional complication, and this is reported in Figures~\ref{Fig.MLDC1B.1.1c_effs}b and~\ref{Fig.MLDC1B.1.1c}. We could correctly establish that the signal was confined to the region spanned by two frequency chunks (corresponding to the number 471 and 472, see Figure~\ref{Fig.MLDC1B.1.1c_effs}b) covering the band $9.942\,\mHz - 9.946\,\mHz$, which is consistent with the true value of the key file, $f_{\mathrm{true}} = 9.94347~\mHz$.
However, running three independent chains with arbitrary starting values and different initialisation seeds produced posterior PDFs which provided ambiguous results.
In Figure~\ref{Fig.MLDC1B.1.1c} we plot the posterior PDFs of three parameters (frequency and sky location), for the 
three chains. We see that for case (a) the MCMC converged on the 
correct point in the parameter space, but for (b) the chain converged onto a secondary maximum of the likelihood function, and
in (c) it did not even converge. This is the effect of a known property of the likelihood function produced by galactic binary signals that is characterised by multiple peaks at a frequency separation $\approx 1\,\mathrm{yr}^{-1}$ \cite{Cornish:2005qw} that our code was not yet sufficiently mature to deal with. Some \emph{ad-hoc} tests could be used in order to discriminate the chains that converged to
the actual place in the parameter space from the other ones, like comparing the recovered SNR or looking at the residuals when the
recovered source was subtracted, but we are focussing on the future applications of the algorithm where it will be necessary to have 
all the chains sampling properly the posterior PDFs in a reasonable number of steps. Thus, work is currently on-going to address this
issue, while retaining the Markovian properties of the chains. It should be noted that the problem of multimodal posterior distributions
is more general, and also present in other sources considered so far in the MLDCs, for example massive-black-hole binary inspirals 
and extreme mass-ratio inspirals.

%%%%%%%%%%%
%%%%%%%%%%%

\section{Conclusions and future work}

We have begun the development of an end-to-end analysis algorithm to identify and study stellar-mass galactic binary systems
in a Bayesian framework. The core of the analysis is based on an application of MCMC techniques. Using a first implementation
of some of the building blocks of this pipeline we have successfully analysed the simplest single-source Challenge 1B.1.1a-b
data sets: the true parameters of the signals all lay within the $95.5 \%$ posterior probability interval, the combined $A$ and $E$ recovered signal-to-noise ratio exceeds $95 \%$ of the optimal one and the correlation between the recovered and the true waveform is above $99 \%$. This performance is competitive with that achieved by other groups who
analysed the same data sets.
Due to the immaturity of the algorithm in dealing with the multiply-peaked structure of the likelihood function for galactic binary signals at higher frequencies, results were not submitted for the  Challenge 1B.1.1c data set.

Work is currently on-going to address the limitations of the present MCMC implementation. In particular, an extension to this algorithm
using a Delayed Rejection MCMC technique has been developed and is undergoing testing and validation. A description of this extended
technique is in preparation. Our long-term goal is to implement a multi-source version using RJMCMC in order to perform model-selection
on the number of sources.

%
%  confident intervals and posterior PDFs. We also have compared our performance with the
%median of other groups'.

%Our algorithm was also able to search for a signal in a wide frequency band ($2$ mHz), however, we found
%some convergence problems analyzing high frequency / low amplitude signals (\eg challenge 1.1c). So,
%our intermediate goal will be to try to fix this kind of problem, in particular the exploration of the islands of
%probability by the chain.
%
%Moreover, a lot of work tuning up the MCMC code can still be done in order to improve its efficiency, \eg instead
%of proposing on linear values of the amplitude, work with $\log \A$; or explore how efficiency gets improved
%when we work with the two uncorrelated parameters $(\A_\plus~,~\A_\cross)$ instead of $(\A~,~\cos \iota)$.
%
%Once we have solved properly the 'one isolated signal problem', our goal will be deal with the real problem of 
%having hundreds of sources overlapped in a narrow frequency band.

%%%%%%%%%%%
%%%%%%%%%%%

\section*{Acknowledgments}
%\acknowledgements 

We are grateful to Alexander Stroeer for sharing his MCMC code \cite{Stroeer:2006ye}, which facilitated the development 
of the software used in this work. MT acknowledges the University of Birmingham for hospitality while this work was
carried out and is grateful for the support of the Spanish Ministerio de Educaci\'on y Ciencia Research Projects 
FPA-2007-60220, HA2007-0042, CSD207-00042 and the Govern de les Illes Balears, Conselleria
d'Economia, Hisenda i Innovaci\'o.  AV and JV acknowledge the support by the UK Science and Technology Facilities Council.

%%%%%%%%%%%
%%%%%%%%%%%

\end{document}